\documentstyle[11pt]{article}
\input epsf

\title{\bf Colour Strings vs. Hard Pomeron in
Perturbative QCD}
\author{M.A.Braun$^{a,b}$ and C.Pajares$^a$\\
$^a$ Dep. of Elementary Particles,\\
Univ. of Santiago de Compostela, 15706,
Santiago de Compostela, Spain, \\
$^b$ Dep. of High Energy physics,
 University of S.Petersburg,\\
198504 S.Petersburg, Russia}
\def\beq{\begin{equation}}
\def\eeq{\end{equation}}
\def\noi{\noindent}
\begin{document}
\maketitle
\medskip
\noi{\bf Abstract.}
Average multiplicities and transverse momenta in AA collisons at high 
energies are studied in the soft and
hard regions, in fusing string  and perturbative QCD scenarios
respectively. Striking similarities are found between the predictions
of the two approaches. Multiplicities per string and average $p_T^2$ are 
found to, respectively, drop and rise with $A$ in a very similar manner, 
so that their product is nearly a constant. In both approaches
total multiplicities grow as $A$, that is, as the number of participants.
The high tail of the $p_T$ distribution in the perturbative QCD scenario
is found to behave $\propto A^{1.1}$.
\vspace{3.5cm}
\section{Introduction}
At present multiparticle production at high energies is
described by different models, which are supposed to be valid in
different intervals of transverse momenta of secondaries.
At small momenta (soft region) one of the most popular and
successful models is the colour string model. In its original
formulation it assumes that in a collision a certain number of
colour strings of definite length in rapidity
are stretched between the colliding partons,
which then independently decay into observed secondaries
 ~\cite{cap,kai}.
The colour string is visualized as a strong colour field
which is succesively broken by creation of quark-antiquark
pairs. A more refined version takes into account not only a finite
length in rapidity but also a finite transverse dimension  of the string.
This inevitably leads to the phenomenon of string fusion and percolation
~\cite{bp1,bp2,bp3}. The colour string model with fusion and percolation
describes quite
satisfactorily multiparticle production in the soft region.
In particular, it predicts that, due to fusion, mutiplicities become
substancially damped, as compared to the independent string picture.
The damping factor $F$ may be related to the so-called percolation
parameter
\beq
\eta=\frac{N\sigma_0}{S},
\eeq
where $\sigma_0$ is the transverse area of the string and $N$ is the
number of strings in the interaction area $S$. As a function of $\eta$
one finds for the damping factor ~\cite{bp3}
\beq
F(\eta)=\sqrt{\frac{1-e^{-\eta}}{\eta}},
\eeq
so that at large $\eta$ multiplicities are damped by $1/\sqrt{\eta}$.

With all this, the colour string model remains mostly phenomenological,
although at the basis of it there lie certain ideas borrowed both from the 
old
Regge phenomenology and  QCD in the limit of large number of colours
(see ~\cite{cap}). Less phenomenological approaches can naturally be
developed in the
hard region where the secondaries are assumed to have large
transverse momenta.
The well known hard scattering picture has been successfully applied to
production of heavy flavour and high-mass Drell-Yan pairs. However this
approach is valid only in the kinematical region appropriate for
the DGLAP evolution, for values of $x$ of the order unity, and the following
ordering in transverse momenta. The region of small $x$ can be reached via
the evolution according to the BFKL equation and its generalization for
nuclei. Both hard approaches suffer from serious drawbacks. The DGLAP
evolution cannot be generalized to several hard collisions in a convincing
manner, since this involves multiparton distributions corresponding
to higher twists. The BFKL approach does not take into account the
running of the coupling. It also
involves small transverse momenta, where
it cannot be valid, and violates unitarity for hadronic scattering.
In this respect the situation is better for scattering off nuclei, where
the small transverse momentum region is strongly damped and unitarity is
automatically fulfilled ~\cite{kov,bra1}. In spite of these difficulties hard
approaches give predictions which are compatible with the experimental
data.

In view of this split between soft and hard regions it is of certain interest
to find a bridge between them. In particular it has long been suspected
that damping of multiplicities predicted by colour string fusion has its
obvious counterpart in the hard region in the form of pomeron fusion
due to pomeron interaction. Note that a literal comparison between the
two approaches is hardly possible. In the colour string picture fusion leads
to appearance of parts of the transverse space with a larger
colour field strength ("strings of higher colour"). As a result, damping
of multiplicities is accompanied by the rise of the average transverse
momentum. In the pomeron picture, at least in the high colour limit,
fusion of pomerons does not lead to new objects. Only the average number of
pomerons may become reduced, and the multiplicities with them.
But one does not naively expect any change in the transverse momentum.
As we shall see in the following  sections, this is fully confirmed in the
simple old-fashioned local supercritical pomeron model, in which also
damping of the multiplicities is found to be much stronger than predicted
by the colour string models with fusion.
The new result reported in this paper is that the perturbative QCD
hard pomeron  approach leads to multiparticle production which
qualitatively fully agrees with the colour string approach with fusion.
Not only damping of the multiplicities turns out to be of the same strength
as in the string picture, but also the average transverse momenta are
found to rise nearly as predicted by the latter model.
So we find an agreement between predictions of these two models, 
pertaining to
completely different (in fact opposite) kinematical regions of 
secondaries,
about certain basic features of multiparticle spectra. 
These results are not fully unexpected. Indeed similar predictions
were found previously ~\cite{LMSB} in the framework of colour glass
condensate ~\cite{LMRV}. Also in all considered approaches scaling in the 
transverse momentum distribution was  observed ~\cite{LMSB,AB,BMP}. 
We consider this
as a strong support for these predictions and thereby for the models.

\section{Generalities. Fusing colour string predictions}
Our basic quantity will be the inclusive cross-section $I_{AB}(y,k)$
to produce a particle  with the transverse momentum
$k$ at rapidity $y$ in a collision of two nuclei with atomic numbers
$A$ and $B$:
\beq
I_{AB}(y,k)=\frac{(2\pi)^2d\sigma}{dyd^2k}.
\eeq
It can be represented as an integral over the impact parameter $b$:
\beq
I_{AB}(y,k)=\int d^2bI_{AB}(y,k,b).
\eeq
To simplify our study we shall concentrate on the inclusive cross-section
at fixed impact parameter $b$.
 We shall also limit ourselves to collision of identical
nuclei $A=B$  and for brevity denote $I_{AA}\equiv I_A$ and so on.
The corresponding multiplicity at fixed rapidity $y$
will be given by
\beq
\mu_{A}^{tot}(y,b)=\frac{1}{\sigma_{A}(b)}
\int\frac{d^2k}{(2\pi)^2}I_{A}(y,k,b),
\eeq
where $\sigma_{A}(b)$ is the total inelastic cross-section for the
collision of two identical nuclei at fixed impact parameter $b$.
To study the effect of string fusion we shall be
interested in the multiplicity per string $\mu_{A}(y,b)$,
given by the ratio of (5) to the number of strings $\nu_{A}(b)$ at
impact parameter $b$
\beq
\mu_{A}(y,b)=\frac{\mu_{A}^{tot}(y,b)}{\nu_{A}(b)}.
\eeq
The latter can be determined by the number of inelastic NN
collisions times the number of strings in a single NN collision
$\nu_{N}$.
For identical nuclei we find
\beq
\nu_{A}(b)=\frac{A^2\sigma_NT_{AA}(b)}{\sigma_{A}(b)}\nu_{N}
\eeq
where $T_{AA}(b)$ is the  nuclear transverse density in the overlap area:
\beq
T_{AA}(b)=\int d^2cT_A(c)T_A(b-c),
\eeq
$\sigma_N$ is the NN total cross-section
and $T_A(b)$ is the standard nuclear profile function for a 
single nucleus $A$ normalized to
unity.
In the ratio (6) the total nucleus-nucleus cross-section $\sigma_A(b)$
cancels;
\beq
\mu_{A}(b)=\frac{\int(d^2k/(2\pi)^2)I_{A}(y,k,b)}
{A^2T_{AA}(b)\sigma_N\nu_{N}}.
\eeq
This point is very important, since it means that we shall  have to
calculate only the inclusive cross-sections for the collision
of two nuclei but not the total cross-sections , which is a problem
of incomparably more complexity.

To simplify the problem still further, we shall consider the
simplest choice of
constant profile function $T_A(b)$ inside a circle of nuclear radius
$R_A=A^{1/3}R_0$. Then also the inclusive cross-section will be
independent of $b$ inside the overlap area. We choose $b=0$
(central collision) when the overlap area coincides with the
nuclear transverse area to find from (9)
\beq
\mu_{A}=A^{-4/3}\frac{\pi R_0^2}{\sigma_N\nu_N}
\int\frac{d^2k}{(2\pi)^2}I_{A}(y,k).
\eeq
Parallel to this we shall study the average transverse momentum
squared, defined by
\beq
<k^2>_{A}=\frac{\int d^2k k^2I_{A}(y,k)}
{\int d^2kI_{A}(y,k)}.
\eeq
Here $b=0$ is implied.
Equations (10) and (11) will be our basic tools in ther following.

We start with the fusing colour strings picture. In it the strength
of fusion is determined by the percolation parameter (1). For central
collisions it is given by
\beq
\eta_{A}=A^{2/3}\frac{\sigma_{N}^2}{\pi^2R_0^4\sigma_{A}(b=0)}\eta_{N},
\eeq
where $\eta_{N}$ is the value of the parameter for $NN$ collisions
at the same energy. Note that the value of $\eta$ depends on the
total inelastic nuclear cross-section for central collisions. We take
it in the optical approximation as
\beq
\sigma_{A}(b=0)=1-e^{-A^{4/3}\sigma_{N}/(\pi R_0^2)}.
\eeq
As stated in the Introduction, the fusing string picture predicts
that multiplicities are damped by the factor (2):
\beq
\mu_{A}=\mu_0F(\eta_{A})
\eeq
where $\mu_0$ is the multiplicity corresponding to a single string.
From this we find
\beq
\frac{\mu_{A}}{\mu_1}=\sqrt{\frac{\eta_1}{\eta_{A}}}
\sqrt{\frac{1-e^{-\eta_{A}}}{1-e^{-\eta_1}}}.
\eeq
This relation describes the $A$-dependence of the
multiplicity. It does not involve the unknown string multiplicity $\mu_0$.
At high string densities and consequently large $\eta$'s it obviously
predicts damping of multiplicities according to
\beq
\mu_{AA}\propto \frac{1}{\sqrt{\eta_A}}\propto A^{-1/3}.
\eeq
Note that, as a result of fusion, the total multiplicity from being  
proportional to a number of inelastic collisions, $\propto A^{4/3}$,
reduces to become proportional to the number of participants
$\propto A$.

It also follows from the fusing string picture that $<k^2>_A$ behaves
inversely to multiplicity. It grows with $\eta$:
\beq
<k^2>_{A}=<k^2>_0\frac{1}{F(\eta_{A})},
\eeq
so that the product $\mu_{A}<k^2>_{A}$ does not change with fusion of 
strings.

\section{Old-fashioned local supercritical pomeron}
In this section we shall compare predictions about the
average multiplicity and transverse momentum which follow from the
colour string model with those from the  old supercritical local
pomeron model.
This will serve as a benchmark for the study in the next section
of analogous predictions following from the non-local perturbative
QCD pomeron.

As shown in ~\cite{cia}, if the nucleus-nucleus interaction is
governed by the exchange of pomerons with the
triple pomeron interactions, the inclusive cross-sections are
given by the convolution of two sets of fan diagrams connecting
the emitted particle to the two nuclei times the vertex for
the emission (Fig. 1).
Explicitly, at a given impact parameter $b$
\beq
I_{AB}(y,k,b)=f(k)\int d^2c\Phi_B(Y-y,b-c)\Phi_A(y,c),
\eeq
where $f(k)$ is the emission vertex, $\Phi_{A,B}$'s are sums of
fan diagrams connected to nuclei A and B and it is assumed that
nucleus A is at rest and the incident nucleus B is at the  overall rapidity
$Y$.

The form (18) characteristic for the old-fashioned local Regge-Gribov theory
immediately tells us that the average transverse momentum does not
depend on $A$ or $B$ and so does not feel fusion of pomerons at all.
It obviously follows from the fact that independent of $A$ or $B$ the
observed particle is emitted from the same single pomeron.
Therefore predictions of the old local pomeron theory for the
transverse momentum dependence do not agree with those from the
fusing colour string model. They rather correspond to models without
fusion, in which indeed $<k^2>$ does not depend on the string density.

Passing to the multiplicities we use the well-known solution for the
fans ~\cite{sch}. Taking $A=B$ and constant  nuclear profile functions
we have for $|b|<R_A$
\beq
\Phi_A(y)=A^{1/3}\frac{g}{R_0^2}\ \frac{e^{\Delta y}}
{1+A^{1/3}\frac{\lambda}{R_0^2\Delta}\left(e^{\Delta y}-1\right)},
\eeq
where $\Delta$ is the pomeron intercept minus one, $\lambda$ (positive) is
the triple pomeron coupling with a minus sign and $g$ is the
pomeron nucleon coupling. Taking $y=Y/2$ (central rapidity) we find the
inclusive cross-section  defined in the previous section as
\beq
I_{A}(y,k)=A^{2/3}
\pi R_0^2f(k)\Big[\Phi_A\left(\frac{Y}{2}\right)\Big]^2
\eeq
The A-dependence of this inclusive cross-section obviously depends on
the energy. At small energies  one may neglect the second term in the
denominator of (19), which actually means that one neglects all non-
linear effects. Then $J_{A}\propto A^{4/3}$ and from (10) one concludes 
that
the multplicity per string does not depend on $A$, as expected.
At large enough $Y$ when one can retain only the exponential term in the
denominator of (19) the inclusive cross-section $I_{A}$ becomes 
proportional to $A^{2/3}$. Then according to (10) the multiplicity per 
string will fall with
$A$ as $1/A^{2/3}$, much faster than predicted by the colour string 
scenario.

So in the end we see that the old-fashioned phenomenological local pomeron
model leads to predictions which do not agree with those from
the fusing colour string  model. The multiplicities fall too fast at
high values of $A$ and the average transverse momentum does not grow at all.
It is remarkable that this situation radically changes with the
perturbative QCD pomeron.

\section{Perturbative QCD pomeron}
The fundamental change introduced by the perturbative QCD approach is
that the pomeron becomes non-local. Its propagation is now governed
by the BFKL equation (see ~\cite{bfkl} for a review). Its interaction is
realized by the triple
pomeron vertex, which is also non-local ~\cite{ver,bar}. Equations
which describe nucleus-nucleus interaction in the perturbative QCD
framework have been obtained in ~\cite{bra2}.
They are quite complicated and
difficult to solve (see ~\cite{bra3} for a partial solution), but
they will be
not needed for our purpose. Knowing that the AGK rules are satisfied
for the diagrams with BFKL pomerons interacting via the triple
pomeron vertex ~\cite{bar} and using arguments of ~\cite{cia} it is
easy to conclude that,
as with the old local pomerons, the inclusive cross-section will again
be given by the convolution of two sums of fan diagrams propagating
from the emitted particle towards the two nuclei. The fundamental
difference will be that the transverse momentum dependence will not be
factorized as in (18) but depend non-trivially on the momenta inside
the two non-local fans.

Taking again $A=B$ and constant nuclear density for $|b|<R_A$  we find
the inclusive cross-section in the perturbative QCD as  ~\cite{bra4}
\beq
I_{A}(y,k)=A^{2/3}\pi R_0^2\frac{8N_c\alpha_s}{k^2}\int 
d^2re^{ikr}[\Delta\Phi_A(Y-y,r)]
[\Delta\Phi_A(y,r)],
\eeq
where $\Delta$ is the two-dimensional Laplacian and $\Phi(y,r)$ is the 
sum 
of all fan diagrams connecting the
pomeron at rapidity $y$ and of the transverse dimension $r$ with the
colliding nuclei, one at rest and the other at rapidity $Y$.
Function $\phi_A(y,r)=\Phi(y,r)/(2\pi r^2)$, in the momentum space,
 satisfies the well-known
non-linear equation ~\cite{kov,bra1,bal}
\beq
\frac{\partial\phi(y,q)}{\partial \bar{y}}=-H\phi(y,q)-\phi^2 (y,q),
\eeq
where $\bar{y}=\bar{\alpha}y$, $\bar{\alpha}=\alpha_sN_c/\pi$,
$\alpha_s$ and $N_c$ are the strong coupling constant and the number
of colours,
respectively, and $H$ is the BFKL Hamiltonian. Eq. (22) has to be solved
with the initial condition at $y=0$ determined by the colour dipole
distribution in the nucleon smeared by the profile function of the
nucleus.

In our study we have taken the initial condition in accordance with
the Golec-Biernat distribution ~\cite{gobi}, duly generalized for the nucleus:
\beq
\phi(0,q)=-\frac{1}{2}a\,{\rm Ei}
\left(-\frac{q^2}{0.3567\, {\rm GeV}^2}\right),
\eeq
with
\beq
a=A^{1/3}\frac{20.8\, {\rm mb}}{\pi R_0^2}.
\eeq
Evolving $\phi(y,q)$ up to values ${\bar y}=3$ we found the
inclusive cross-section (21) at center rapidity  for energies corresponding
to the overall rapidity  $Y=\bar{Y}/\bar{\alpha}$.
with $\bar{Y}=6$. Taking $\alpha_s=0.2$
this gives $Y\sim 31$, which is far beyond the present possibilities.
The overall cutoffs for
integration momenta in Eq.(22) were taken according to
$0.3.10^{-8}$ GeV/c  $< q < 0.3.10^{+16}$ GeV/c.

The found inclusive cross-sections are illustrated in Figs 2-6.
To see how the form of the distribution changes with energy,
we present in Fig. 2  the distributions for $A=1$ and $y=Y/2$
normalized to unity and multiplied 
by $k^2$ to
exclude the trivial $1/k^2$ dependence present in (21),
\beq
J_{1}(y,k)=k^2I_{1}(y,k)/\int \frac{d^2k}{(2\pi)^2}I_{1}(y,k)
\eeq
at different
energies corresponding to $\bar{Y}=1,3,6$.
One observes how, with the growth of energy, the distributions are shifted
towards higher values of $k$.

In Figs. 3-6  we illustrate the $A$-dependence
showing ratios
\beq
R^{col}_{A}=
\frac{I_{A}(y,k)}{A^{4/3}I_{1}(y,k)}
\eeq 
and
\beq
R^{part}_{A}=
\frac{I_{A}(y,k)}{AI_{1}(y,k)}
\eeq 
with inclusive cross-sections scaled by {\it the number of collisions},
$\propto A^{4/3}$, or by {\it the number of participants} $\propto A$,
at $y=Y/2$ and $Y=1,3$ and 6.
One clearly sees that whereas at relatively small momenta the inclusive 
cross-sections are proprtional to $A$, that is to {\it the number of 
participants}, at
larger momenta they grow with $A$ faster, however  noticeably slowlier
than the number of collisions, approximately as $A^{1.1}$.
The interval of momenta for which $I_A\propto A$ is growing with energy,
so that one may conjecture that at infinite energies all the spectrum 
will be proportional to $A$.

Passing to the determination of both multiplicities and
average transverse momenta one has to observe certain care because
of the properties of the perturbative QCD solution in the leading
approximation embodied in Eq. (22).  As follows from (21) the inclusive
cross-section blows up at $k^2\to 0$ independent of the rapidity $y$.
So the corresponding total multiplicity diverges logarithmically.
However, the physical sense has only emission of jets
with high enough transverse momenta. Thus we restricted ourselves
to produced jets with $k>k_{min}$. For $k_{min}$ we chose two possibilities:
$k_{min}=0.3$ and 1.0  GeV/c. Our conclusions turned
out to be practically independent of this choice.
In the following we discuss the results with $k_{min}=0.3$ GeV/c.
As to the average transverse momentum, the calculated $\phi(y,q)$
fall very slowly with $q$, so that $<k^2>$ clearly diverges.
Even the calculation of $<|k|>$, which converges, encounters
certain difficulties at highest $Y$ due to reduced precision and
influence of overall cutoffs. So we found our values of average
$k^2$ squared as $<|k|>^2$.

Due to unreasonably high value of the BFKL intercept
$\Delta=\bar{\alpha}\,4\ln 2$,
both the multiplicities and average transverse momenta
grow very fast with $Y$
and reach unreasonably high values $\mu\sim 10^6$ and
$<|k|>\sim 10^4$ GeV/c at $\bar{Y}=6$. However we are not interested
in the $Y$ dependence but rather in the $A$-dependence, since in the
fusing string scenario the energy dependence is introduced on the
phenomenological grounds. 

To a very good precision, at high energies corresponding
to scaled rapidities $\bar{Y}>2$ the total 
multiplicities are found to be 
proportional to $A$, that is to {\it the number of participants}.

To compare with the string scenario we turn to multiplicities per string
$\mu_A$.
In Tables 1 and 2 the ratios
\beq
r_\mu(A)=\frac{\mu_A}{\mu_1},\ \ \
r_k(A)=\left(\frac{<|k|>_A}{<|k|>_1}\right)^2
\eeq
at given overall scaled rapidities $\bar{Y}=1,...6$ .   Table 3 shows the
product $r_\mu r_k$ which is unity  in the fusion strings model.
These results  were obtained with the low $k_{min}$.
\begin{table}[h,t]

\begin{center}
\caption{Ratios $r_\mu(A)=\frac{\mu_A}{\mu_1}$}
\vskip 2mm

\begin{tabular}{|c|c|c|c|c|c|c|}\hline
    &   &    &        &      & & \\
A &$\bar{Y}$=1    & 2 &3 &4 &5&6\\
    &    &   &        &      &    &     \\\hline
  8&  0.695&  0.611&  0.585&  0.572&  0.565&  0.560 \\
 27&  0.531&  0.439&  0.412&  0.400&  0.392&  0.387 \\
 64&  0.429&  0.341&  0.317&  0.306&  0.299&  0.294 \\
125&  0.358&  0.278&  0.256&  0.246&  0.240&  0.236 \\
216&  0.306&  0.233&  0.214&  0.205&  0.200&  0.196 \\\hline
\end{tabular}
\end{center}
\vskip 2mm
\label{table1}
\end{table}
\begin{table}[h,t]

\begin{center}
\caption{Ratios $r_k(A)=\left(\frac{<|k|>_A}{<|k|>_1}\right)^2$}
\vskip 2mm

\begin{tabular}{|c|c|c|c|c|c|c|}\hline
    &   &    &        &      & & \\
A &$\bar{Y}$=1    & 2 &3 &4 &5&6\\
    &   &    &        &      &    &     \\\hline
  8&  1.305&  1.606&  1.718&  1.745&  1.713&  1.651 \\
 27&  1.559&  2.123&  2.336&  2.378&  2.310&  2.214 \\
 64&  1.810&  2.643&  2.922&  3.001&  2.839&  2.685 \\
125&  2.025&  3.117&  3.508&  3.568&  3.319&  3.105 \\
216&  2.226&  3.575&  4.010&  4.070&  3.761&  3.488 \\\hline
\end{tabular}
\end{center}
\vskip 2mm
\label{table2}
\end{table}
\begin{table}[h,t]

\begin{center}
\caption{Products $r_\mu(A)r_k(A)$}
\vskip 2mm

\begin{tabular}{|c|c|c|c|c|c|c|}\hline
    &   &    &        &      & & \\
A &$\bar{Y}$=1    & 2 &3 &4 &5&6\\
    &   &    &        &      &    &     \\\hline
  8&  0.907&  0.982&  1.004&  0.998&  0.967&  0.924 \\
 27&  0.828&  0.933&  0.963&  0.950&  0.906&  0.857 \\
 64&  0.776&  0.902&  0.926&  0.917&  0.848&  0.790 \\
125&  0.725&  0.866&  0.898&  0.878&  0.797&  0.732 \\
216&  0.681&  0.834&  0.858&  0.834&  0.751&  0.684 \\\hline
\end{tabular}
\end{center}
\vskip 2mm
\label{table3}
\end{table}

Already a superficial study of these results shows their striking
similarity with the predictions based on the fusing string picture.
In particular the products $r_\mu(A)r_k(A)$ result nearly universal
and close to the value unity predicted by the latter. A certain drop
of this product towards higher $A$ and $\bar{Y}$ may to be related
to the neglected far tail of the momentum distribution at super-high
values of $k$, which is absolutely irrelevant for the multiplicity
but can give some contribution to $<|k|>$.

A more detailed comparison can be performed using Eq. (15) derived from
the string picture. We fitted the  better known ratios $r_\mu(A)$
according to this equation  using $\eta_1$ as an adjustable parameter
and taking for $\sigma_{N}(s)$ the experimental data well-reproduced by
\beq
\sigma_{N}=38.3+0.545\ln^2\left(\frac{s}{122\, {\rm GeV}^2}\right)
\eeq
The values of $\eta_1$ which give the best least-square fit at
overall scaled rapidities $\bar{Y}=1,2,...6$ are respectively
\[\eta_1=0.267,\ \ 0.678,\ \ 0.931,\ \ 1.201,\ \ 1.348,\ \ 1.433\]
With these $\eta_1$'s we reproduce the data for $r_\mu(A)$ from
Table 1 by Eq. (15) with the average relative error which goes from
the maximal 2.8\% at $\bar{Y}=1$ down to minimal 0.98\% at $\bar{Y}=6$.
Using the adjusted value of $\eta_1$ at $\bar{Y}=1$ and assuming that
at this comparatively low energy the number of strings in the $NN$
collision is exactly 2, we could determine the effective string radius
corresponding to the pomeron picture to be $r_0=0.32$ fm. This value is
astonishingly close to standard values used in the fusing string
calculations.
If we assume that this string radius  is fixed independent of energy
then we can find the effective average number of strings in a NN collision
at higher energies. We find at $\bar{Y}=2,3,...6$ respectively
\[\nu_{N}=3.72,\ \ 4.31,\ \ 3.71,\ \ 2.81,\ \ 2.11\]
It is interesting that at accessible energies (up to $\bar{Y}=3$)
the number of strings monotonously grows in accordance with the standard
expectations, although noticeably slowlier. This may be interpreted
as a signal of string fusion already in NN collisions. At still higher
energies this phenomenon seems to become much stronger so that the
effective average of strings begins to fall.

So to conclude the predictions from the perturbative QCD pomeron approach
seem to fully agree with those from the fusing string picture.

\section{Conclusions}
We have compared predictions for multiplicities and average transverse
momentum which follow from the semi-phenomenological
fusing colour string picture for the soft domain with those which
follow from the pomeron approach, both phenomenological and perturbatively
derived from QCD.

The old-fashioned pomeron approach with triple pomeron interaction
leads to results which disagree with
the  colour string models both with or without fusion.
The average transverse momentum in this approach is independent of $A$,
contrary to predictions of the colour model with fusion.
On the other hand, the multiplicity per string falls with $A$, in contradiction
with the predictions of the models without fusion. In fact the only way
to reconciliate the two models is to assume the eikonal form for
the multiple pomeron exchange without the triple pomeron interaction.
Such a model gives prediction equivalent to the independent string
picture (without fusion). In fact this is the dynamics tacitly assumed
in the original form of the string model ~\cite{cap,kai}

The perturbative QCD pomeron gives results which are  in remarkable
agreement with the string model with fusion. The behaviour
of the multiplicities and average transverse momentum are in good
agreement not only qualitatively but also quantitatively. Moreover the
effective string radius extracted from these results turns out to be
in agreement with the standardly assumed value in fusing colour string
calculations.  This overall agreement may appear to be astonishing
in view of very different dynamical pictures put in the basis of the
two approaches and also quite different domains of their applicability:
soft for the string picture and hard for the pomeron picture. However, on
second thought, one may come to the conclusion that the dymanical difference
between the two approaches is not so unbridgeable. Two phenomenons are
playing the leading role in both approaches. One is fusion of exchanged
elemental objects, strings in one picture and pomerons in the other.
This explains damping of multiplicities per one initial elemental object.
Second phenomenon is the rise of average transverse momentum with this
fusion. It is generated by formation of strings of higher tension (colour)
in the string scenario. In the pomeron model this rise occurs due
the growth of the so-called saturation momentum, which shifts the momentum
distribution to higher momenta with $A$ (and $Y$). Due to this shift
non-linear effects in Eq. (22) in some sense reproduce formation of
strings of higher colour in the string model.

The discovered similarity in predictions between the perturbative QCD
pomeron and fusing string model indicates that the dynamics
of strong interaction does not radically change 
when passing from the soft to very hard region, in spite of
the change in its microscopic content, from strings to partons.
It also leaves certain hopes that
these predictions are well founded in spite of all known limitations
in validity and applicability of these models.

\section{Acknowledgements}
The authors are deeply indebted to N.Armesto for a constructive
discussion and valuable comments. This work has been supported by
a NATO Grant PST.CLG.980287, a contract FPA2002-01161 from CICYT of 
Spain and a contract PGIDIT03 from Galicia.  

\newpage
\begin{figure}[ht]
\unitlength=1mm
\special{em:linewidth 0.4pt}
\linethickness{0.4pt}
\begin{picture}(93.00,129.33)
\put(30.33,129.33){\line(4,-5){30.00}}
\put(91.67,129.33){\line(-5,-6){31.00}}
\put(60.33,127.33){\line(-1,-1){16.00}}
\put(60.33,91.33){\line(0,-1){24.00}}
\put(60.33,67.67){\line(-5,-6){28.67}}
\put(60.33,67.33){\line(1,-1){32.67}}
\put(76.00,51.33){\line(-2,-3){11.67}}
\put(45.33,80.33){\line(1,0){30.33}}
\end{picture}
\caption{A typical diagram for the inclusive cross-section in nucleus-nucleus
collisions.}
\label {Fig1}
\end{figure}
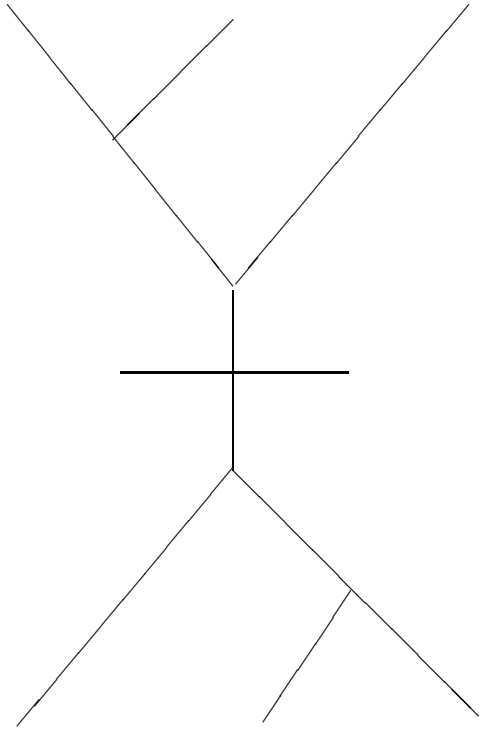
\begin{figure}[ht]
\epsfxsize 4in
\centerline{\epsfbox{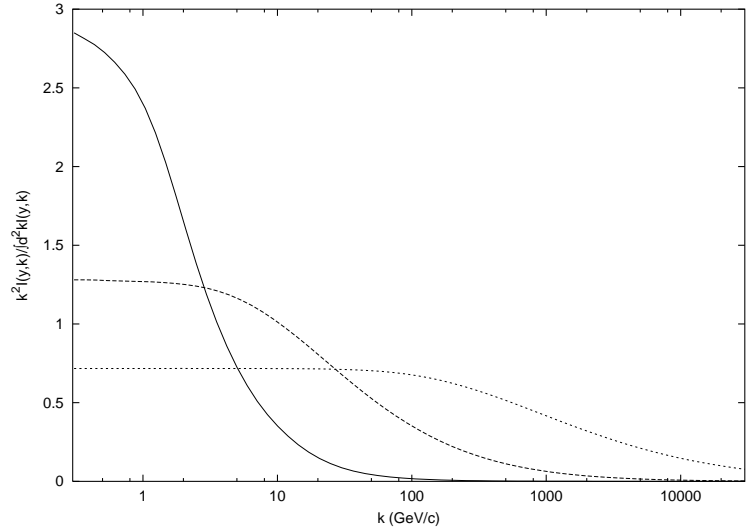}}
\caption{Normalized distributions $J_1(y,k)$ (Eq. (25))
at $y=\frac{Y}{2}$ .
Curves from top to bottom at small $k$ correspond to scaled overall
rapidities $\bar{Y}=1,3,6$.}
\label{Fig2}
\end{figure}
\begin{figure}[ht]
\centerline{\epsfbox{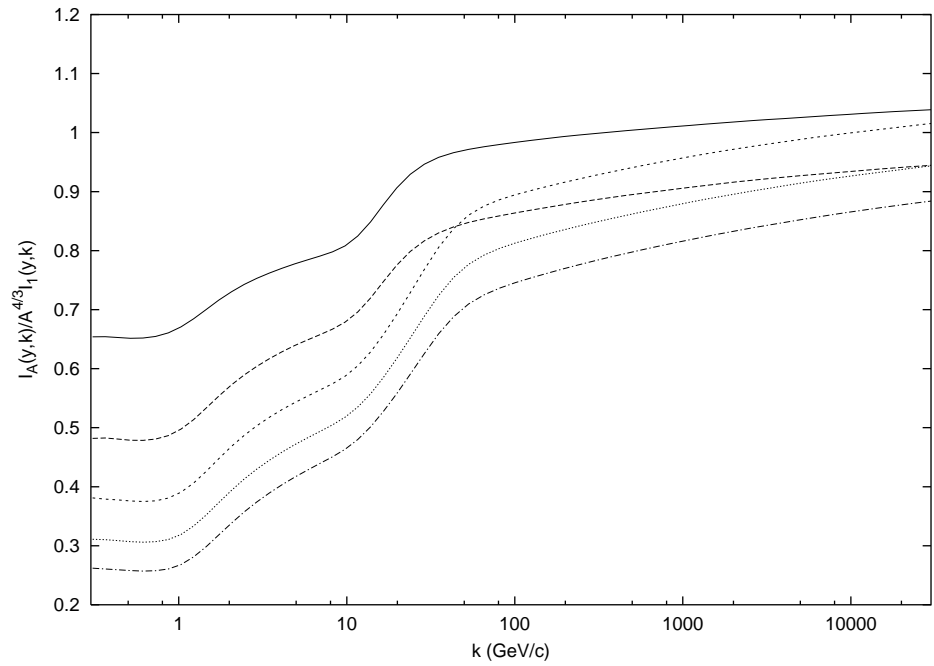}}
\caption{A-dependence of momentum distributions,
scaled with the number of collisions, at $\bar{Y}=1$.
Curves from top to bottom  show ratios $I_A(y,k)/A^{4/3}I_1(y,k)$
at center rapidity ($y=\frac{Y}{2}$) for $A=8,27,64,125$ and 216.}
\label{Fig3}
\end{figure}
\begin{figure}[ht]
\epsfxsize 4in
\centerline{\epsfbox{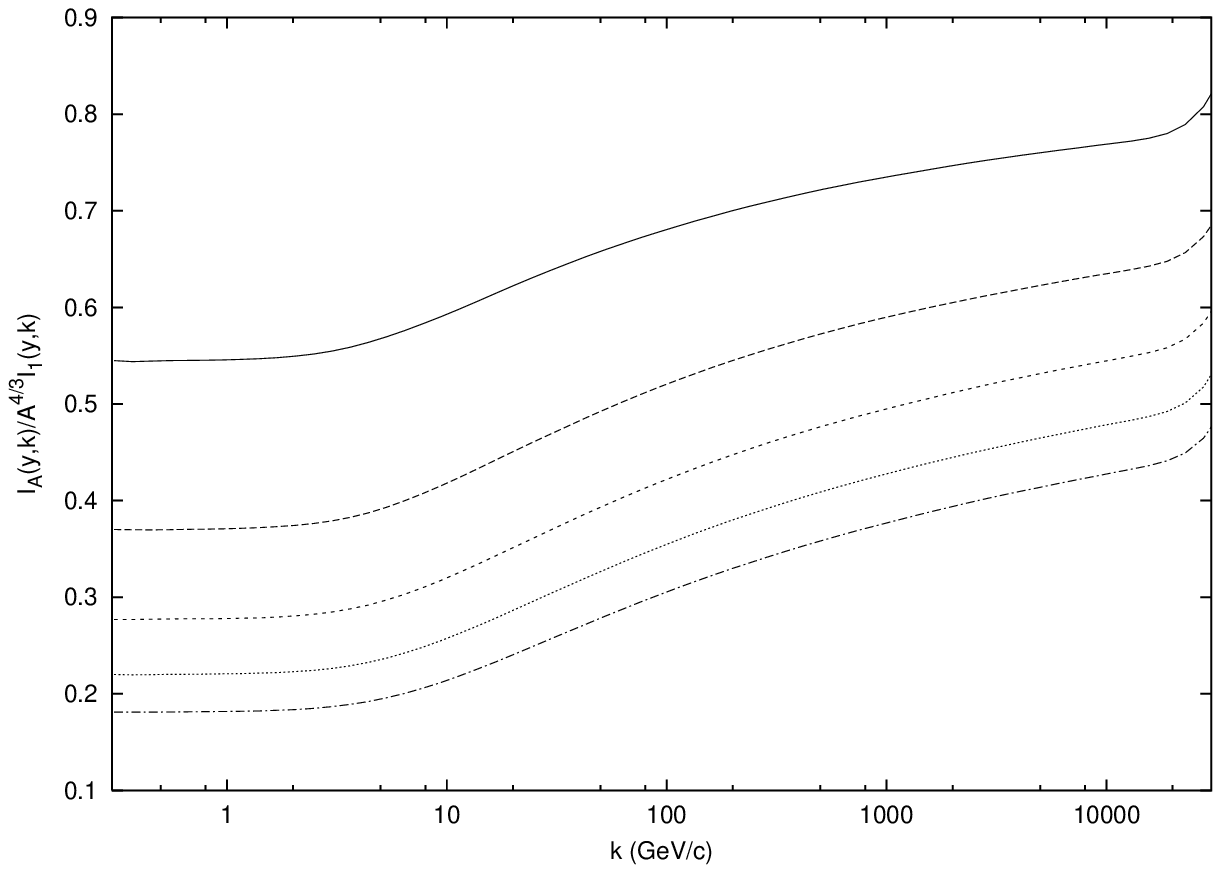}}
\caption{Same as Fig. 3 for $\bar{Y}=3$.}
\label{Fig4}
\end{figure}
\begin{figure}[ht]
\epsfxsize 4in
\centerline{\epsfbox{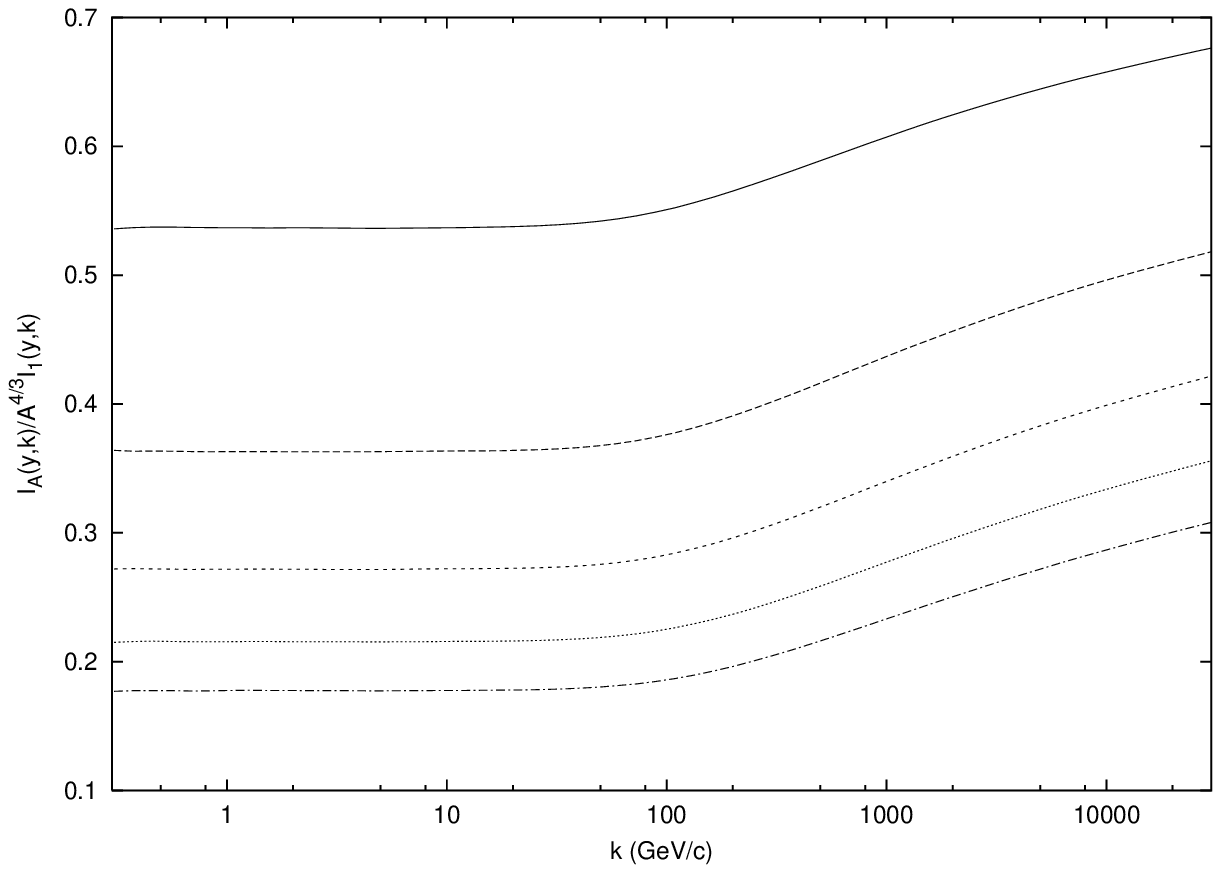}}
\caption{Same as Fig. 3 for $\bar{Y}=6$.}
\label{Fig5}
\end{figure}
\begin{figure}[ht]
\epsfxsize 4in
\centerline{\epsfbox{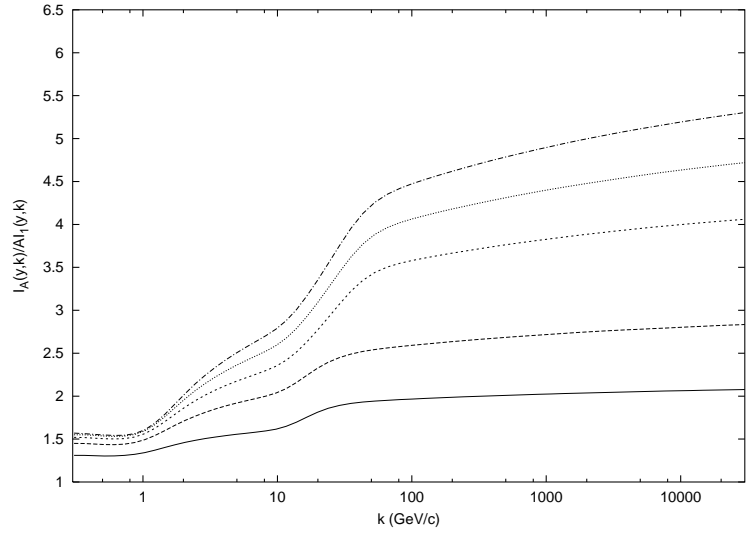}}
\caption{A-dependence of momentum distributions,
scaled with the number of participants, at $\bar{Y}=1$.
Curves from bottom to top show ratios $I_A(y,k)/AI_1(y,k)$
at center rapidity ($y=\frac{Y}{2}$) for $A=8,27,64,125$ and 216.}
\label{Fig6}
\end{figure}
\begin{figure}[ht]
\epsfxsize 4in
\centerline{\epsfbox{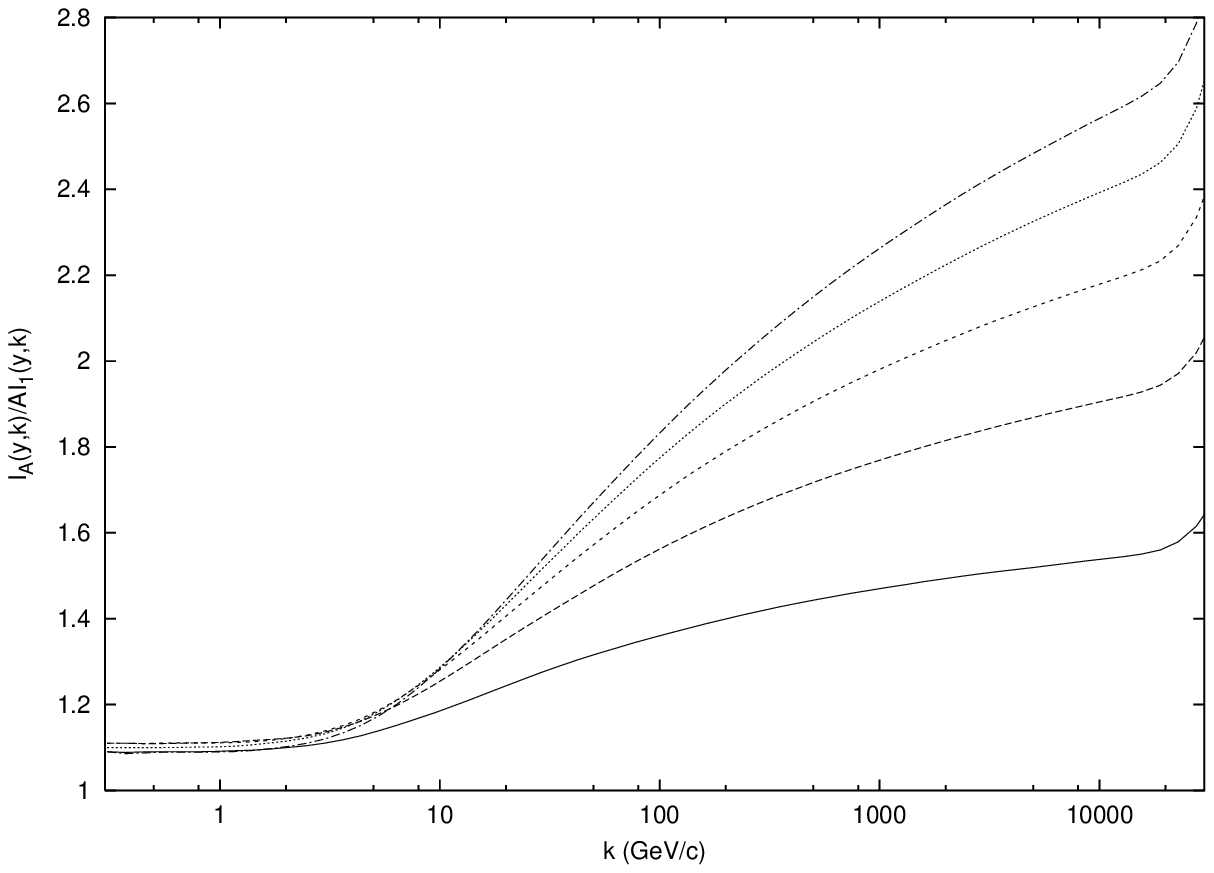}}
\caption{Same as Fig. 6 for $\bar{Y}=3$.}
\label{Fig7}
\end{figure}
\begin{figure}[ht]
\epsfxsize 4in
\centerline{\epsfbox{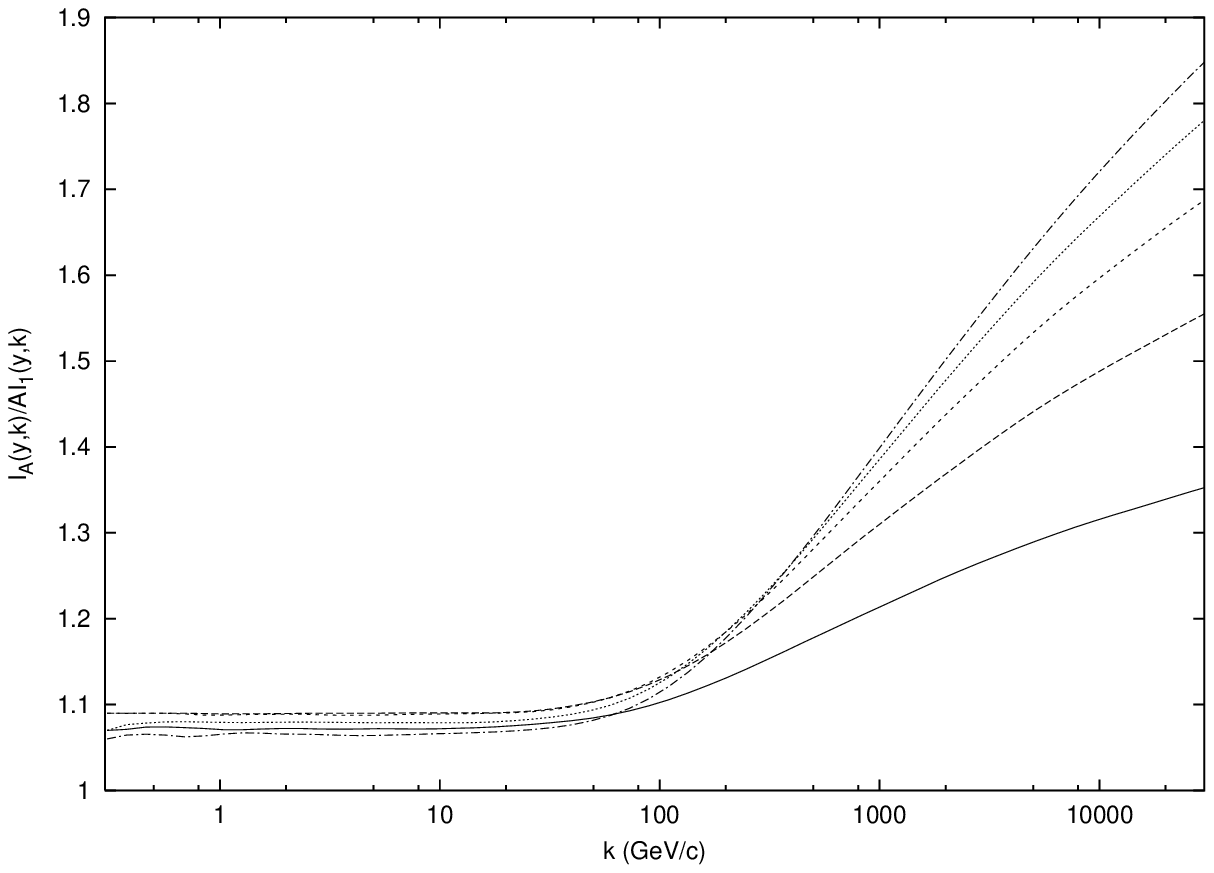}}
\caption{Same as Fig. 6 for $\bar{Y}=6$.}
\label{Fig8}
\end{figure}
\end{document}